\documentclass[aps,prc,twocolumn,showpacs,superscriptaddress,footinbib]{revtex4-1}

\usepackage{graphicx}
\usepackage{dcolumn}
\usepackage{bm}
\usepackage{enumerate}
\usepackage{amsmath}

\usepackage[normalem]{ulem}  

\begin{document}


\title{Double shape quantum phase transitions in the SU3-IBM: new $\gamma$-soft phase and the shape phase transition from the new $\gamma$-soft phase to the prolate shape}

\author{De-hao Zhao}
\affiliation{School of Physics, Liaoning University, Shenyang 110036, People's Republic of China}

\author{Xiao-shen Kang}
\email{kangxiaoshen@lnu.edu.cn}
\affiliation{School of Physics, Liaoning University, Shenyang 110036, People's Republic of China}

\author{Li Gong}
\email{gongli@lnu.edu.cn}
\affiliation{School of Physics, Liaoning University, Shenyang 110036, People's Republic of China}

\author{Ze-yu Yin}
\affiliation{College of Physics, Tonghua Normal University, Tonghua 134000, People's Republic of China}

\author{Tao Wang}
\email{suiyueqiaoqiao@163.com}
\affiliation{College of Physics, Tonghua Normal University, Tonghua 134000, People's Republic of China}

\date{\today}

\begin{abstract}
Shape quantum phase transition is an important topic in nuclear structure. In this paper, we begin to study the shape quantum phase transition in the SU3-IBM. In this new proposed model, spherical-like spectra was found to resolve the spherical nucleus puzzle, which is a new $\gamma$-soft rotational mode. In this paper, the shape phase transition along the new $\gamma$-soft line is first discussed, and then the neighbouring case at the prolate side is also studied. We find that double shape phase transitions occur along a single parameter path. The new $\gamma$-softness is really a shape phase and the shape phase transition from the new $\gamma$-soft phase to the prolate shape is found. The experimental support is also found and $^{108}$Pd may be the critical nucleus.
\end{abstract}

\maketitle

\section{Introduction}

Phase transitions are widely found in nature \cite{Stanley71,Sachdev11}. A common example is that, under standard atmospheric pressure, when the temperature rises the ice becomes water and then water vapor. If the atmospheric pressure is raised to a certain level, the water and water vapor cannot be distinguished. In the field of atomic nuclei, nuclear shape can change when the number of the protons or neutrons varies, and shape quantum phase transition can occur \cite{Casten06,Casten07,Bonatsos09,Casten09,Jolie09,Casten10,Jolos21,Fortunato21,Cejnar21,Jolie00,Cejnar03,Iachello04,Wang08}. Since this control parameter is discrete and finite, it becomes even more interesting to identify these phase transitions \cite{Casten99}.

50 years ago, the interacting boson model (IBM) was proposed by Arima and Iachello, which is an influential algebraic model for describing the collective behaviors of nucleons. In the simplest case, only the $s$ ($L=0$) and $d$ ($L=2$) bosons are considered, and the Hamiltonian has the U(6) symmetry. There are four dynamical symmetry limits (see Fig. 1(a) left): (1) the U(5) symmetry limit can present the spherical shape and its vibration, (2) the SU(3) symmetry limit can describe the prolate shape and its rotation, (3) the O(6) symmetry limit can describe the $\gamma$-soft rotation, and (4) the $\overline{\textrm{SU(3)}}$ symmetry limit can present the oblate shape and its rotation.

This simple model can describe the shape phase transitions between the spherical shape to various quadrupole deformations or among different deformed shapes (see Fig. 1(a) left) \cite{Casten06,Casten07,Bonatsos09,Casten09,Jolie09,Casten10,Jolos21,Fortunato21,Cejnar21,Jolie00,Cejnar03,Iachello04,Wang08}. In these studies, along a single parameter path, the shape of the nucleus changes only from one to another.  After 2000, an important class of shape phase transition has attracted attentions and created controversies, which is the prolate-oblate shape phase transition \cite{Jolie01,Bonatsos24}. In previous IBM, the prolate-oblate shape phase transition is described via changing from the SU(3) symmetry limit to the $\overline{\textrm{SU(3)}}$ symmetry limit, and the O(6) symmetry limit is just the first-order phase transitional critical point \cite{Jolie01}, which implies that the O(6) $\gamma$-softness is not a shape phase. In this description, the spectra of the prolate and oblate shapes are the same, and it is not found in realistic nuclei. In \cite{Jolie03}, for realistic nuclei in the Hf-Hg region, the energy ratio $E_{4/2}=E_{4_{1}^{+}}/E_{2_{1}^{+}}$  of the $4_{1}^{+}$ and $2_{1}^{+}$ states is 3.33 for the prolate shape while 2.55 for the oblate shape ($E_{4/2}$ is not related to the boson number $N$). Thus this mirror symmetry appears not to exist.

Around 2020, an extension of the interacting boson model with SU(3) higher-order interactions (SU3-IBM) was proposed by one of the authors (T. Wang) \cite{Wang20,Wang22}, which incorporates the idea of previous IBM and the SU(3) correspondence of the rigid triaxial shape \cite{Isacker85,Draayer87,Draayer881,Isacker00,Kota20}. In this new model, the role of the SU(3) symmetry is raised to a new level, dominating all the quadrupole deformations of nuclei (see Fig. 1(a) right). It contains only the U(5) symmetry limit and the SU(3) symmetry limit. In the SU(3) symmetry limit, higher-order interactions are needed. The SU(3) second-order Casimir operator $-\hat{C}_{2}[SU(3)]$ can present the prolate shape while the SU(3) third-order Casimir operator $\hat{C}_{3}[SU(3)]$ can describe the oblate shape, which is very different from the $\overline{\textrm{SU(3)}}$ description in previous IBM. The two interactions, together with the square of the SU(3) second-order Casimir operator $\hat{C}_{2}^{2}[SU(3)]$, can describe any rigid triaxial shapes.

The SU3-IBM can be used to explain the B(E2) anomaly \cite{Wang20,zhang14,Zhang22,Wangtao,Zhang24,Pan24,Zhang25,Zhang252,Cheng25,Li25,Tie25}, the Cd puzzle \cite{Wang22,Wang25,WangPd}, the prolate-oblate asymmetric shape phase transition \cite{Fortunato11,Zhang12,Wang23}, the $\gamma$-softness in $^{196}$Pt at a better level \cite{WangPt,ZhouPt}, the E(5)-like spectra in $^{82}$Kr \cite{Zhou23}, the rigid triaxiality in $^{166}$Er \cite{ZhouEr}, and the boson number odd-even effect in $^{196-204}$Hg \cite{WangHg}. Together these results reveal that, the SU3-IBM can better describe the collective behaviors in nuclei.

\begin{figure}[tbh]
\includegraphics[scale=0.33]{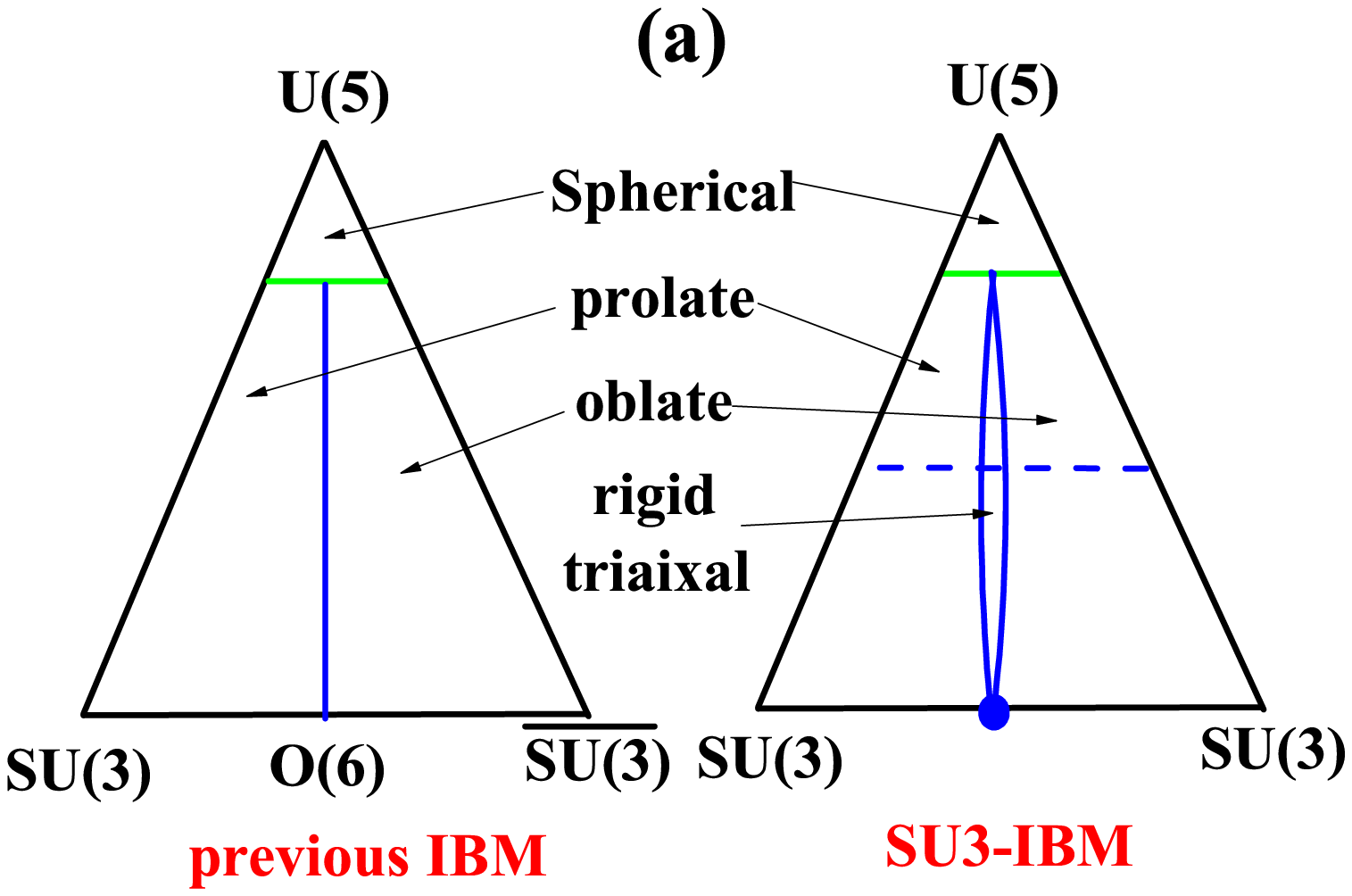}
\includegraphics[scale=0.33]{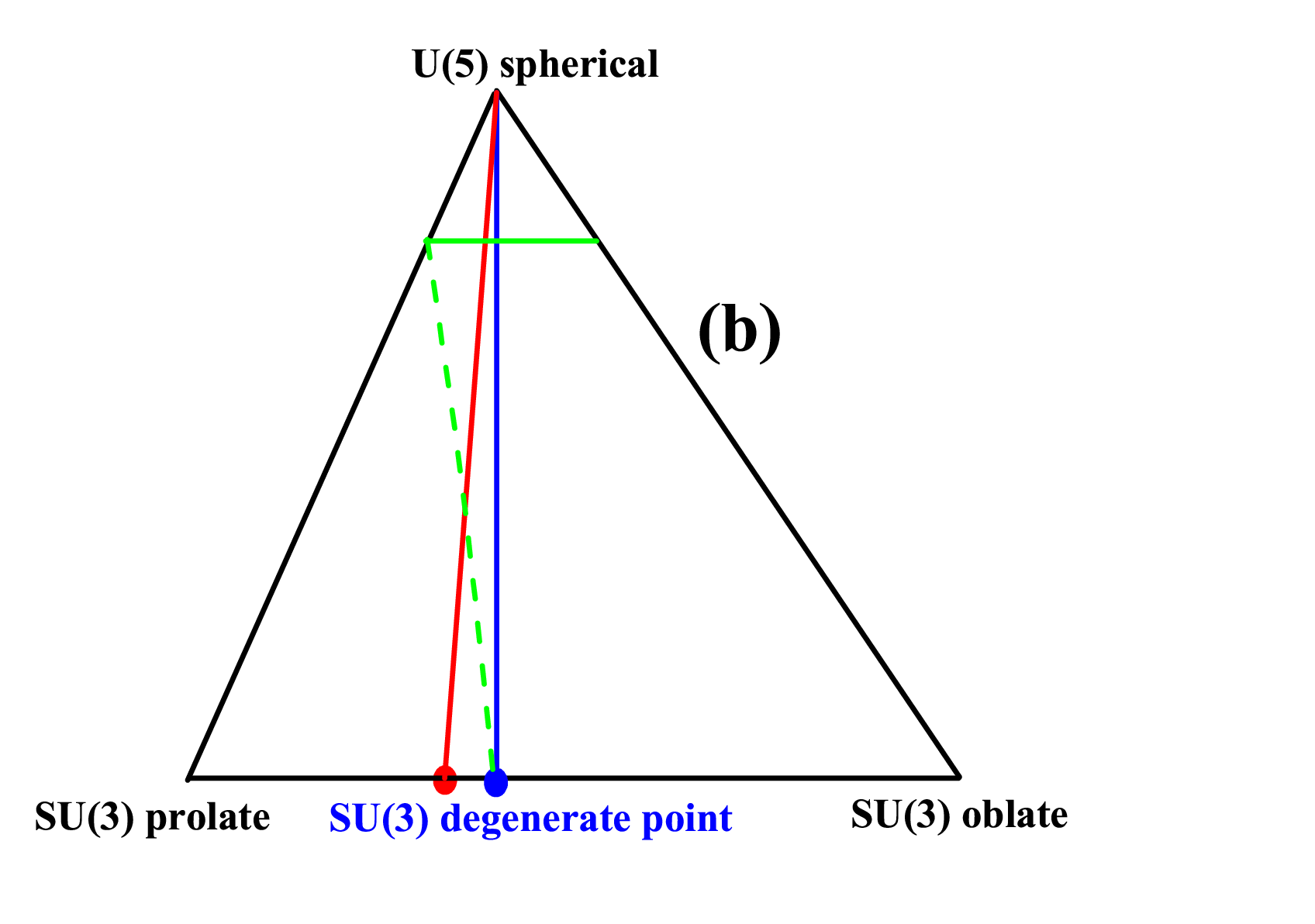}
\caption{(a) left represents the phase diagram of previous IBM while (a) right represents the phase diagram of the $\hat{H}$ in the SU3-IBM. In (b) the real blue and real red lines are two evolutional paths discussed in this paper. Along the real red line, the double shape phase transitions can occur.}
\end{figure}

Thus investigating the shape phase transition in the SU3-IBM is also important. In the SU3-IBM, for resolving the Cd puzzle in Cd nuclei and other nuclei previously thought to be spherical \cite{Garrett18}, the spherical-like nucleus was proposed \cite{Wang20}, which is a new collective excitation and has been verified in realistic nuclei recently \cite{Wang25,WangPd}. This new shape was not mentioned by previous nuclear theories. Moreover the prolate-oblate asymmetric shape phase transition in the Hf-Hg region can be better described by the SU3-IBM \cite{Wang23}. These new studies imply that realistic nuclei can show more rich and complex shape phase transition behaviors, which may be described by the SU3-IBM. So it becomes even more important to study the characteristics of shape phase transitions in the SU3-IBM to help us understand the shape phase transition of actual nuclei.

This is the first paper on this topic. In the SU3-IBM, the existence of the spherical-like spectra is the most important. We study the shape phase transitions from this new collectivity. This has been first discussed in \cite{Wang22} for small boson number $N=7$. In this paper, we discuss them with $N=60$ for the ground state and $N=35$ for the excited states. This evolutional path can be represented by the real blue line in Fig. 1(b) from the U(5) symmetry limit to the SU(3) degenerate point. In this paper, the nearby evolutional path (denoted by the real red line) is also discussed, and we find that, along the real red lines, double shape quantum phase transitions can occur. The first is from the spherical shape to the new $\gamma$-soft rotation, and the second is from the new $\gamma$-soft mode to the prolate shape. Here the new $\gamma$-softness is really a shape phase, which is different from the O(6) critical $\gamma$-softness.  These results look more realistic and very meaningful.

\section{Hamiltonian}

The simplest Hamiltonian for describing the shape phase transition related to the new $\gamma$-softness in the SU3-IBM is as follows \cite{Wang22,Wang23}
\begin{equation}
\hat{H}=c[(1-\eta)\hat{n}_{d}+\eta(-\frac{\hat{C}_{2}[\textrm{SU(3)}]}{2N}+\kappa\frac{\hat{C}_{3}[\textrm{SU(3)}]}{2N^{2}})],
\end{equation}
here $\eta$, $\kappa$ are two controlling parameters and $c$ is the energy scale parameter. $0\leq \eta \leq 1$ and $\kappa \geq 0$. If $\eta=0$, it presents the spherical shape. If $\eta=1$ and $\kappa=0$, it describe the prolate shape. The two cases are the same as the ones in previous IBM \cite{Wang08}. If $\eta=1$ and $\kappa$ varies, this Hamiltonian describes the prolate-oblate shape phase transition \cite{Zhang12}, which is a finite-$N$ effect.

Fig. 1(a) left shows the phase diagram of previous IBM in the large-$N$ limit. Above the real green line, the spherical shape exists, and under the real green line, the deformed shapes exist. The deformed region is divided by the real blue line which is part of the connected line between the U(5) symmetry limit to the O(6) symmetry limit. The left part of the blue line presents the prolate shape while the right part presents the oblate shape. The blue line is the critical line between the prolate and oblate shapes. The crossover point of the green line and the blue line is the triple point with the spherical, prolate and oblate shapes \cite{Casten02,Warner02}. Obviously, the O(6) $\gamma$-softness in not a shape phase. Up to two-body interactions, the IBM can not describe the rigid triaxial shapes \cite{Chen1981}.

\begin{figure}[tbh]
\includegraphics[scale=0.33]{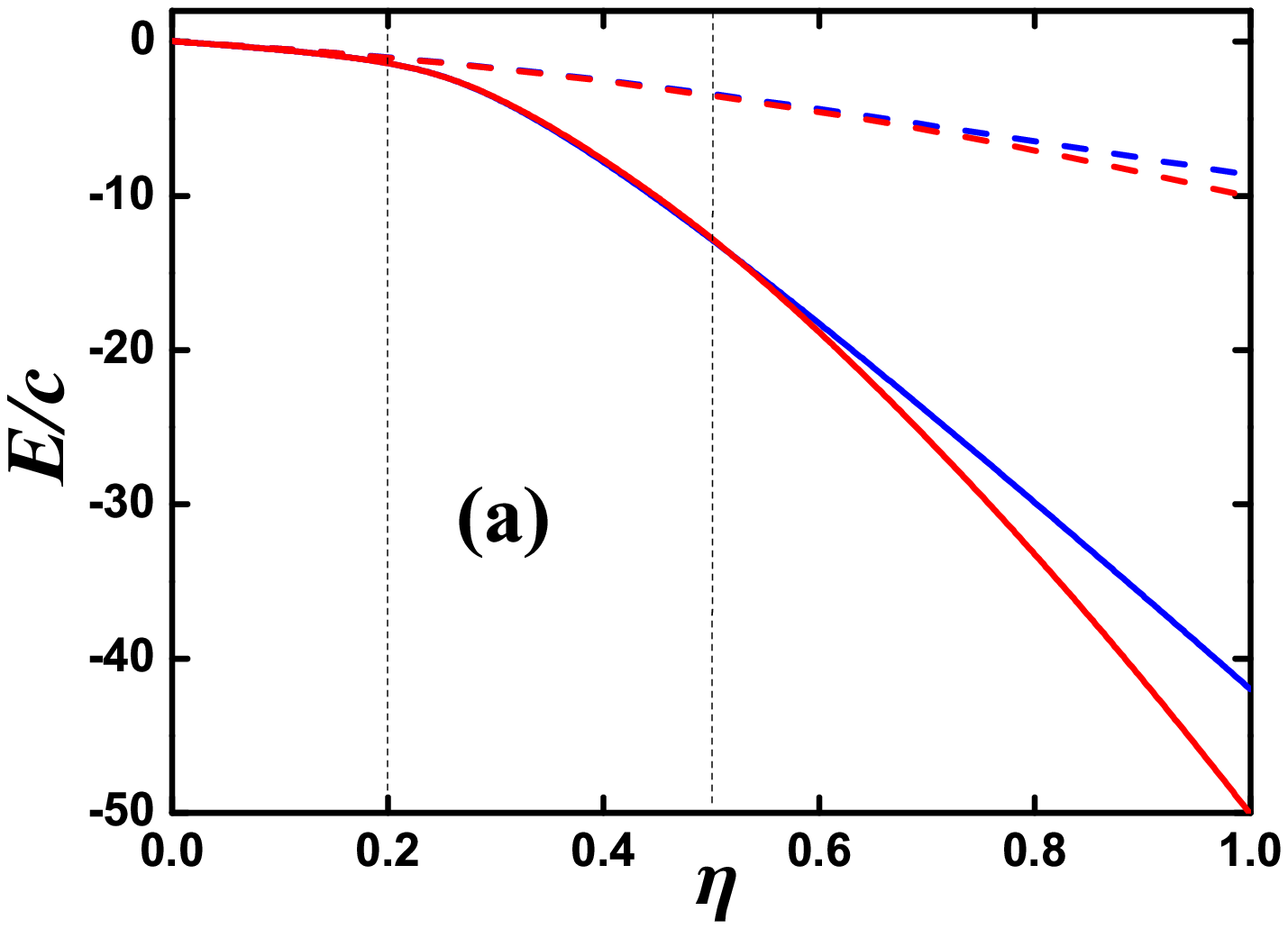}
\includegraphics[scale=0.33]{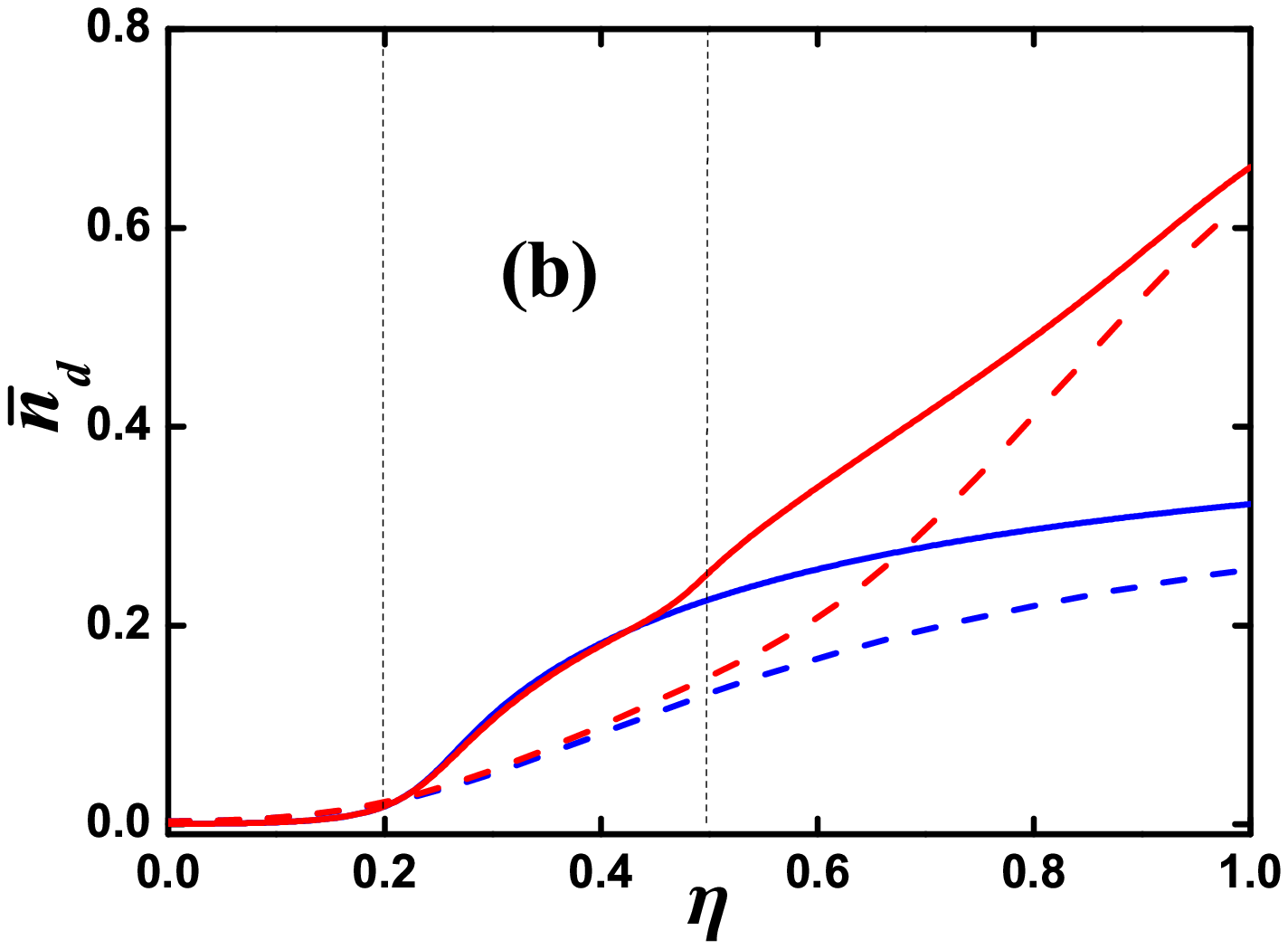}
\caption{(a) The energy evolution of the ground state of $\hat{H}$ along the real blue (red) line in Fig. 1(b) for $N=10$ (dashed blue (red) line) and for $N=60$ (real blue (red) line). (a) The $\bar{n}_{d}$ evolution of the ground state of $\hat{H}$ along the real blue (red) line in Fig. 1(b) for $N=10$ (dashed blue (red) line) and for $N=60$ (real blue (red) line).}
\end{figure}

The phase diagram of $\hat{H}$ in the large-$N$ limit was first discussed in \cite{Fortunato11}, see Fig. 1(a) right (or see Fig. 14 in \cite{Fortunato11}). The key difference between the SU3-IBM and previous IBM is to use the SU(3) third-order Casimir operator $\hat{C}_{3}[SU(3)]$ instead of the $\overline{\textrm{SU(3)}}$ symmetry limit to describe the oblate nuclei. Above the real green line, the spherical shape exists. Under the real green line, the deformed shapes exist. For the deformed region, the phase diagram becomes more complicated than the ones in previous IBM \cite{Wang08}. From the SU(3) prolate to the SU(3) oblate, there exists a SU(3) degenerate point (the blue point). At the SU(3) prolate side of the degenerate point, the shape of the ground state is always the prolate shape, and at the SU(3) oblate side of the degenerate point, it is always oblate. Thus across the SU(3) degenerate point, the shape changes abruptly. This finite-$N$ first-order shape phase phase transition was studied in \cite{Zhang12}. Connected with the SU(3) degenerate point, in the middle of the deformed region, there exists a shallow narrow region with rigid triaxial shapes, see the region between the two blue lines in Fig. 1(a) right. At the left side of the narrow region, the prolate shape exists, while at the right side, the oblate shape exists. Thus along the dashed blue line in Fig. 1(a) right, the shape changes from the prolate to the oblate via a narrow region with rigid triaxiality \cite{Fortunato11}. Although the rigid triaxial region is narrow and shallow, the rigid triaxial shape really exists. The crossover point between the green line and the two blue lines is a fourfold point.

In \cite{Wang22}, one of the authors (T. Wang) found an important result. For finite-$N$, the rigid triaxial shape becomes the new $\gamma$-soft rotational mode. This new $\gamma$-softness is a shape phase and not a critical phenomenon. Along the dashed blue line in Fig. 1(a) right, the shape changes from the prolate to the new $\gamma$-soft, and then to the oblate, which was used to describe the prolate-oblate asymmetric shape phase transition in the Hf-Hg region \cite{Wang23}.

In \cite{Wang22}, the real blue line in Fig. 1(b) is a critical line between the prolate shape and the new $\gamma$-soft rotation in the deformed region, along which the $4_{1}^{+}$, $2_{2}^{+}$ states are degenerate and $6_{1}^{+}$, $4_{2}^{+}$, $3_{1}^{+}$ and $2_{3}^{+}$ states are also degenerate. However for large-$N$, this critical line is actually a curve and curves to the left. Thus for large-$N$, the critical line can not be described by the real blue line \cite{WangPt}. At the right side of the degenerate line, there exists another line, along which the $4_{1}^{+}$, $2_{2}^{+}$ states are also degenerate \cite{WangPt}. Between the two degenerate lines of the $4_{1}^{+}$, $2_{2}^{+}$ states, it was supposed that the new $\gamma$-softness exists, however in this paper it is shown that this new $\gamma$-soft region may be larger, which is unexpected. And importantly we find that double shape quantum phase transition can be observed along a single parameter path.

In this paper, the shape phase transition along the real blue line is studied, and not stress the degenerate line, which is difficult to discuss. The SU(3) degenerate point is at $\kappa_{0}=\frac{3N}{2N+3}$ and the red point is at $0.9\kappa_{0}$. Through previous analysis, the red point presents the prolate shape. This is very interesting. For the real red line from the U(5) symmetry limit to the red point, intuitively, the shape changes from the spherical shape to the prolate. Through later numerical calculations, we find that this shape transition is not direct but through the new $\gamma$-softness region, and the double shape quantum phase transition can occur. The right part of the real blue line is not discussed in this paper, and will be studied in next paper for investigating the scope of the new $\gamma$-soft region \cite{Zhao2}. 

It is important to emphasize here why the O(6) $\gamma$-softness in previous IBM and actual $\gamma$-soft nuclei do not match. In actual nuclei, there exists many nuclei in the $\gamma$-soft region, such as Os, Pt, Xe, Ba nuclei \cite{ensdf}, so it is hard to believe that this is just the O(6)-softness in previous IBM, a critical point of shape phase transition. In the SU3-IBM, such a conceptual conflict does not exist. We believe that the shape phase transition given by the SU3-IBM is an accurate description of the realistic shape phase transitions in nuclei. This point has been preliminarily confirmed by the prolate-oblate asymmetric shape phase transition in the Hf-Hg region \cite{Wang23}.

For understanding the B(E2) anomaly, the $B(E2)$ values are necessary. The $E2$ operator is defined as
\begin{equation}
\hat{T}(E2)=q\hat{Q},
\end{equation}
where $q$ is the boson effective charge. The evolutions of $B(E2; 2_{1}^{+}\rightarrow 0_{1}^{+})$, $B(E2; 4_{1}^{+}\rightarrow 2_{1}^{+})$, $B(E2; 2_{2}^{+}\rightarrow 2_{1}^{+})$, $B(E2; 0_{2}^{+}\rightarrow 2_{1}^{+})$ values are discussed.

\begin{figure}[tbh]
\includegraphics[scale=0.33]{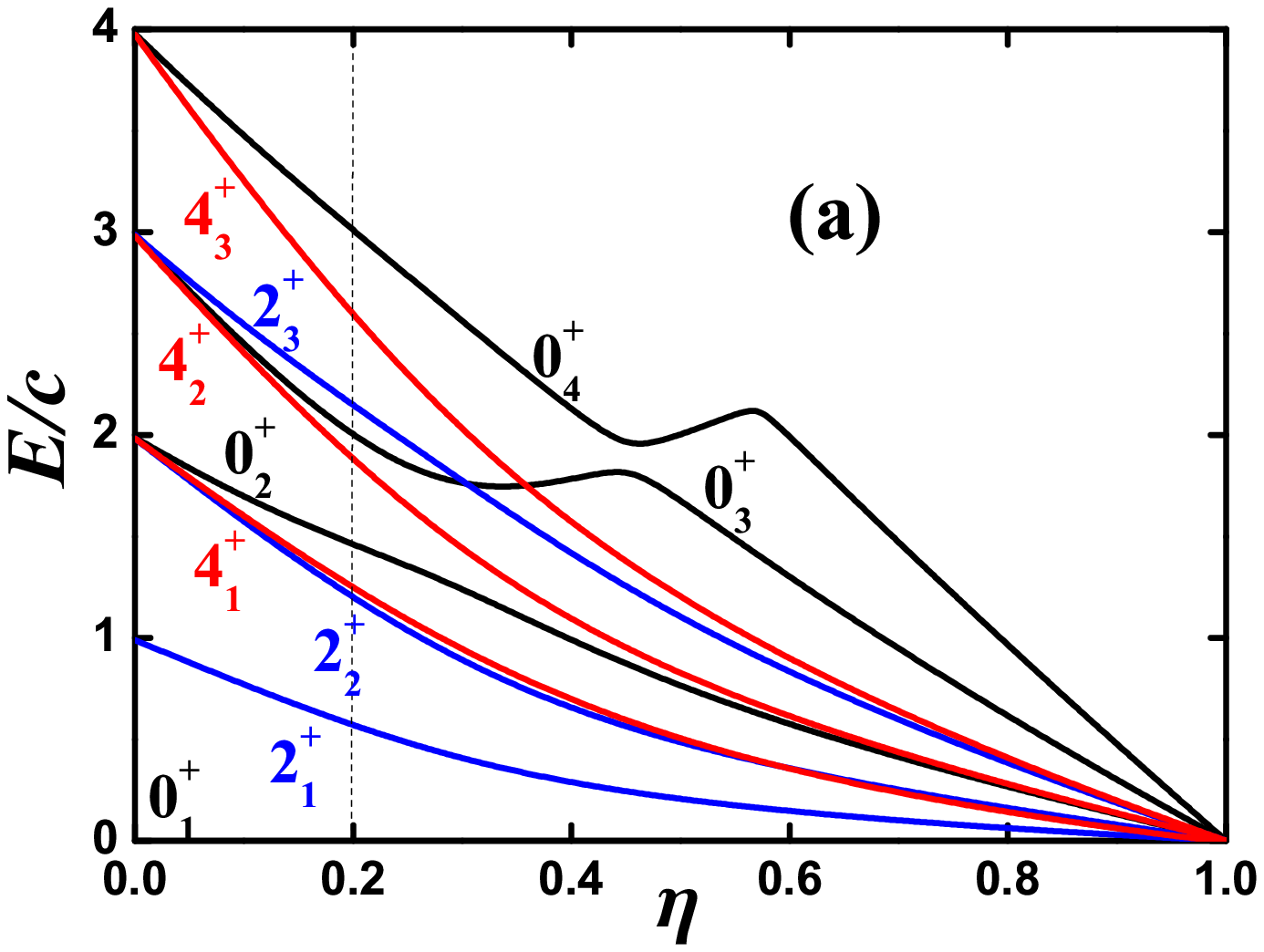}
\includegraphics[scale=0.33]{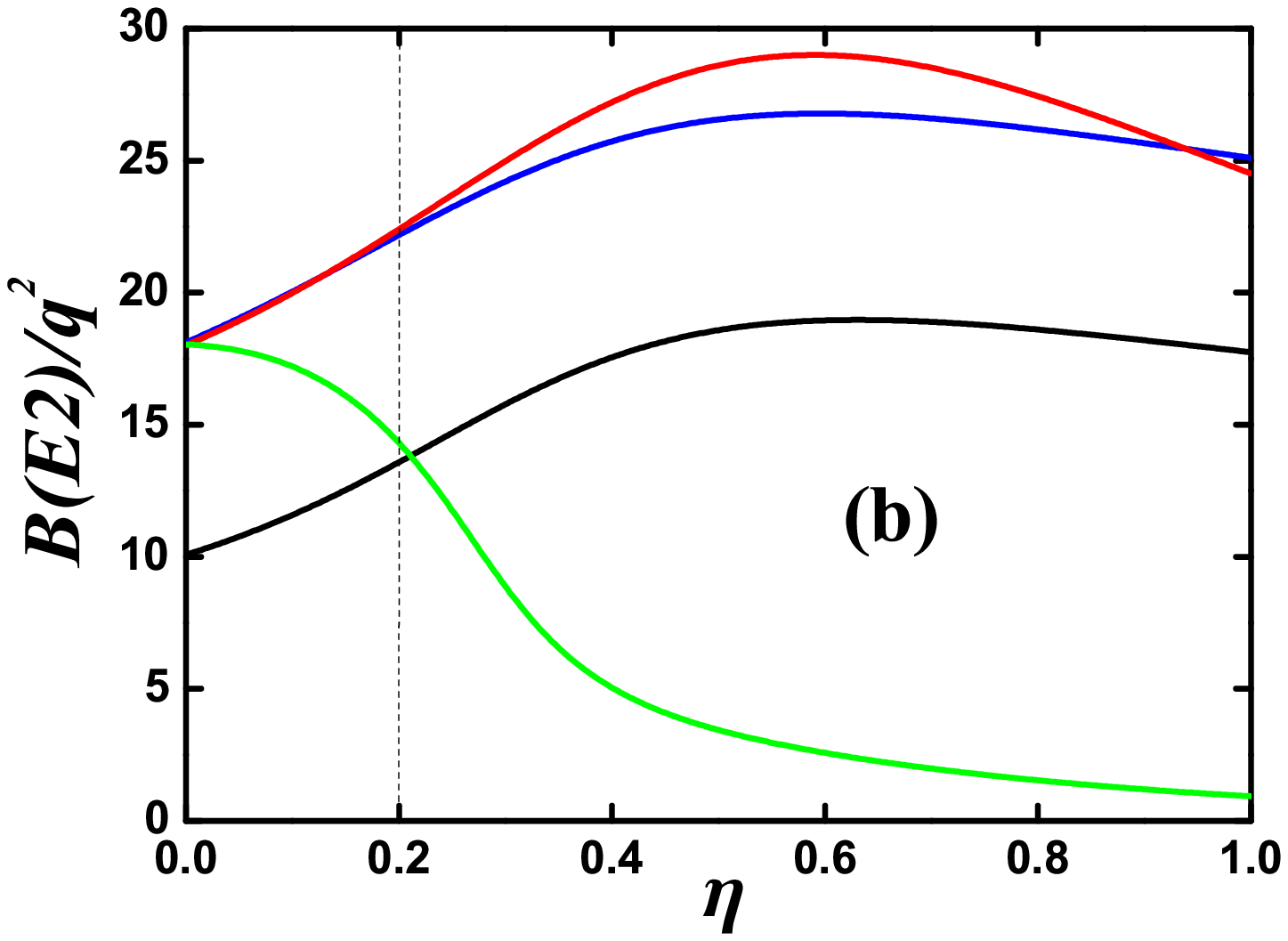}
\caption{(a) The evolutional behaviors of the partial low-lying levels as a function of $\eta$ for $N=10$ along the real blue line in Fig. 1(b); (b) The evolutional behaviors of the $B(E2; 2_{1}^{+}\rightarrow 0_{1}^{+})$ (black line), $B(E2; 4_{1}^{+}\rightarrow 2_{1}^{+})$ (blue line), $B(E2; 2_{2}^{+}\rightarrow 2_{1}^{+})$ (red line), $B(E2; 0_{2}^{+}\rightarrow 2_{1}^{+})$ (green line) along the real blue line in Fig. 1(b).}
\end{figure}

\section{Double shape quantum phase transition}

\begin{figure}[tbh]
\includegraphics[scale=0.33]{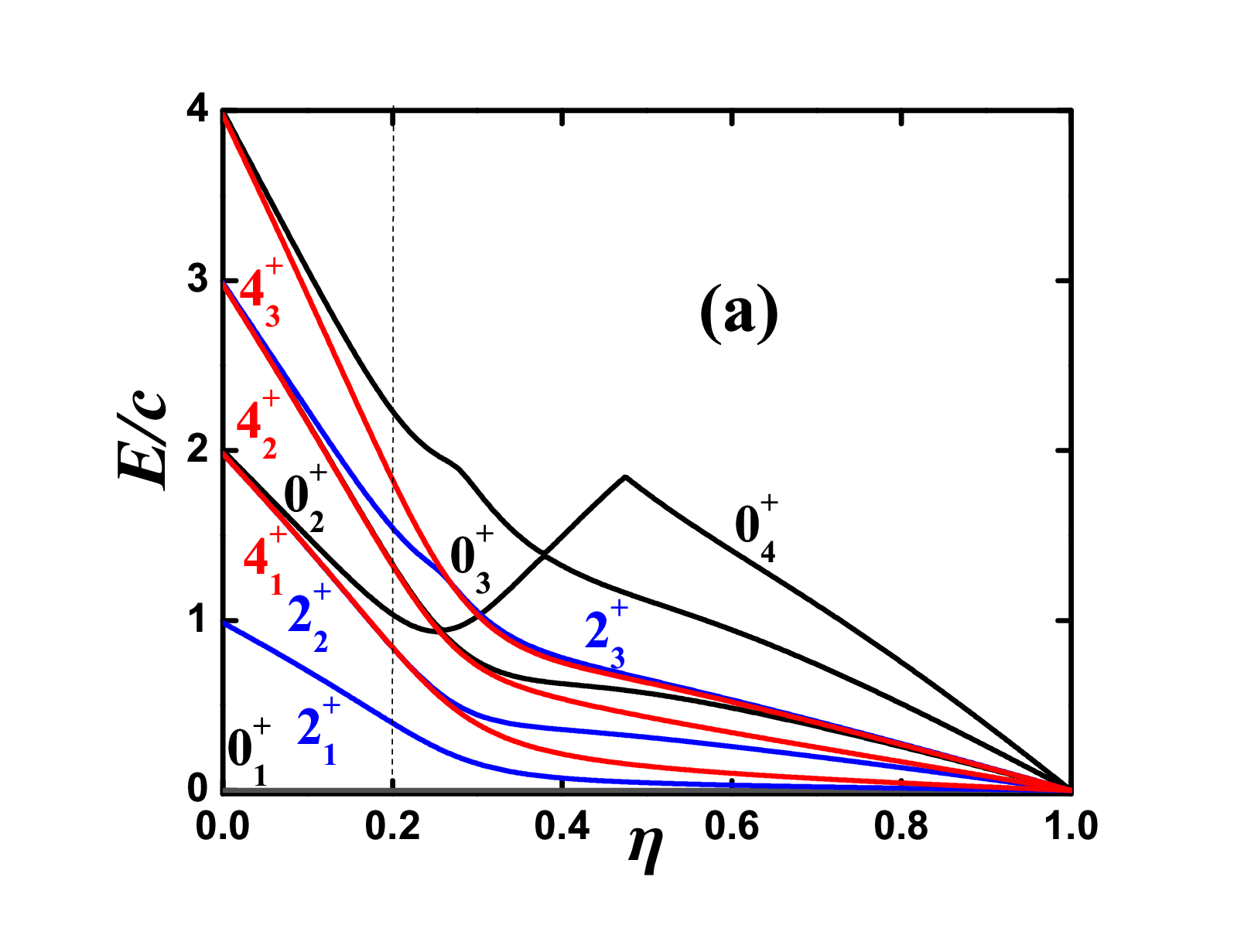}
\includegraphics[scale=0.33]{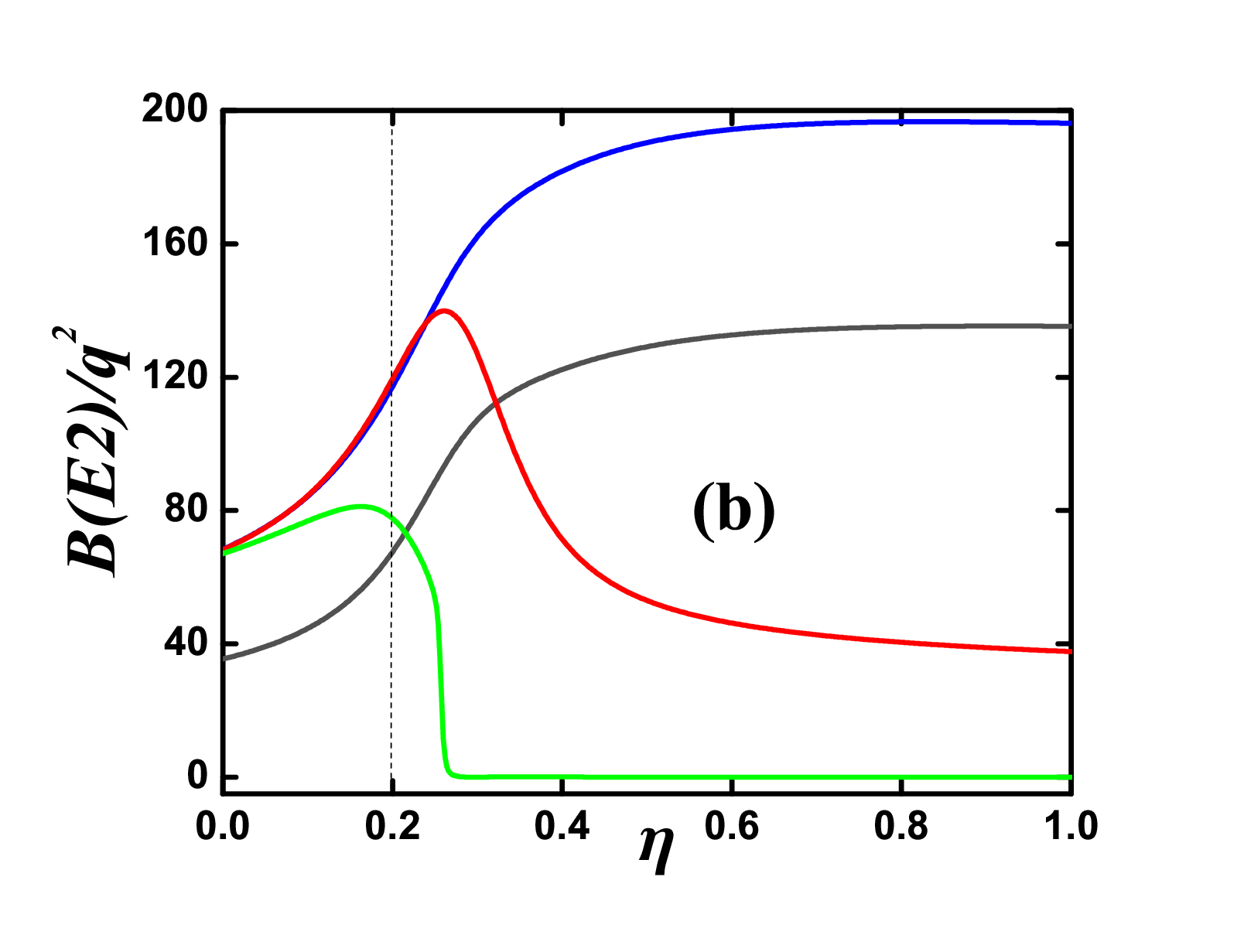}
\caption{(a) The evolutional behaviors of the partial low-lying levels as a function of $\eta$ for $N=35$ along the real blue line in Fig. 1(b); (b) The evolutional behaviors of the $B(E2; 2_{1}^{+}\rightarrow 0_{1}^{+})$ (black line), $B(E2; 4_{1}^{+}\rightarrow 2_{1}^{+})$ (blue line), $B(E2; 2_{2}^{+}\rightarrow 2_{1}^{+})$ (red line), $B(E2; 0_{2}^{+}\rightarrow 2_{1}^{+})$ (green line) along the real blue line in Fig. 1(b).}
\end{figure}

Shape quantum phase transition is first manifested by the energy evolution of the ground state. This is not an observable quantity, but very useful for understanding the shape quantum phase transition. Fig. 2(a) shows the energy evolution of the ground state of $\hat{H}$ along the real blue line in Fig. 1(b) for $N=10$ (dashed blue line) and for $N=60$ (real blue line). Clearly, around $\eta=0.2$ (denoted by the left dashed line), the shape phase transition from the spherical to the new $\gamma$-soft occurs. Between $\eta=0.2$ and $\eta=1$, the new $\gamma$-softness exists. It should be noticed that the SU(3) degenerate point is not $\gamma$-soft.

\begin{figure}[tbh]
\includegraphics[scale=0.33]{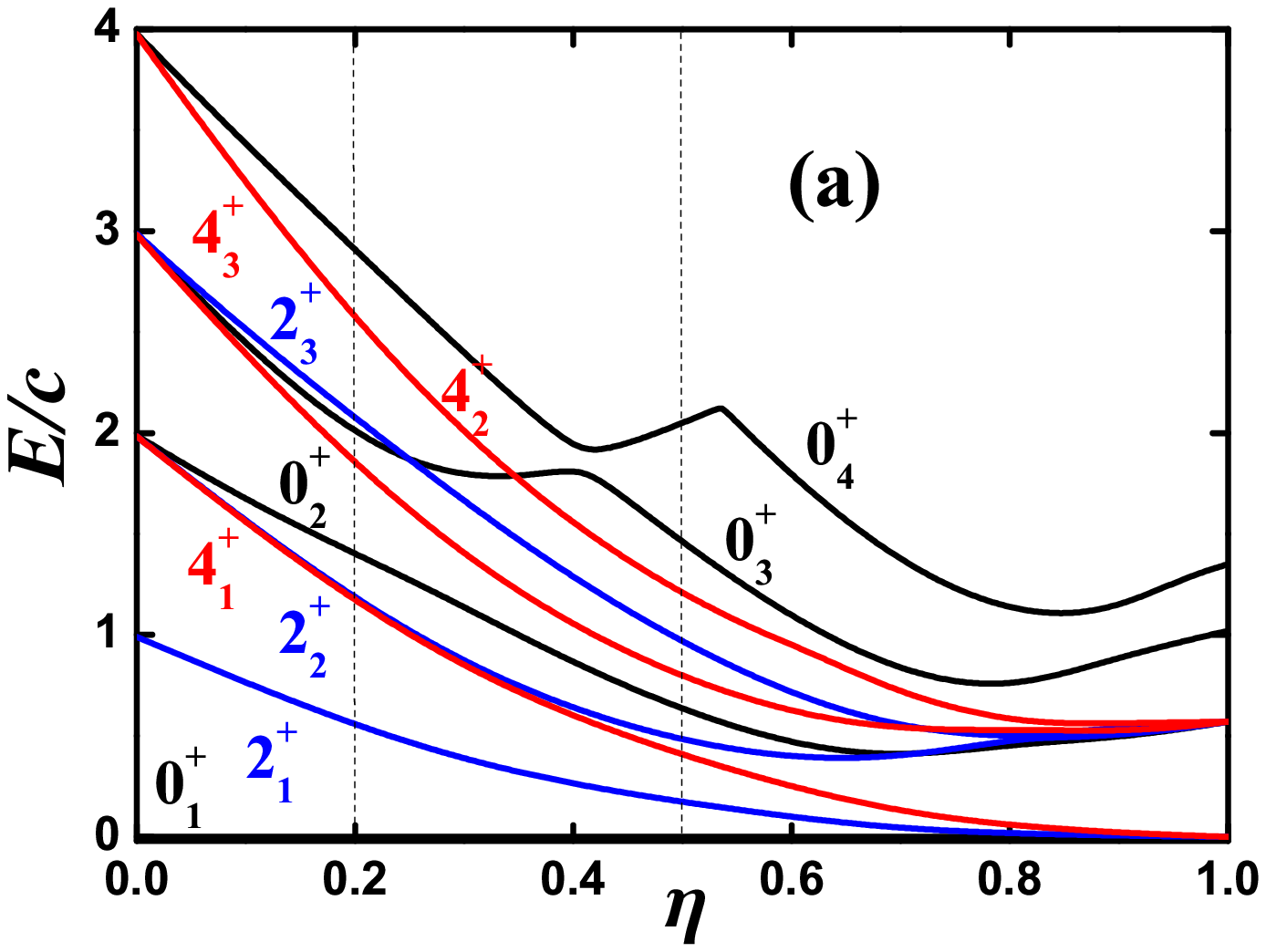}
\includegraphics[scale=0.33]{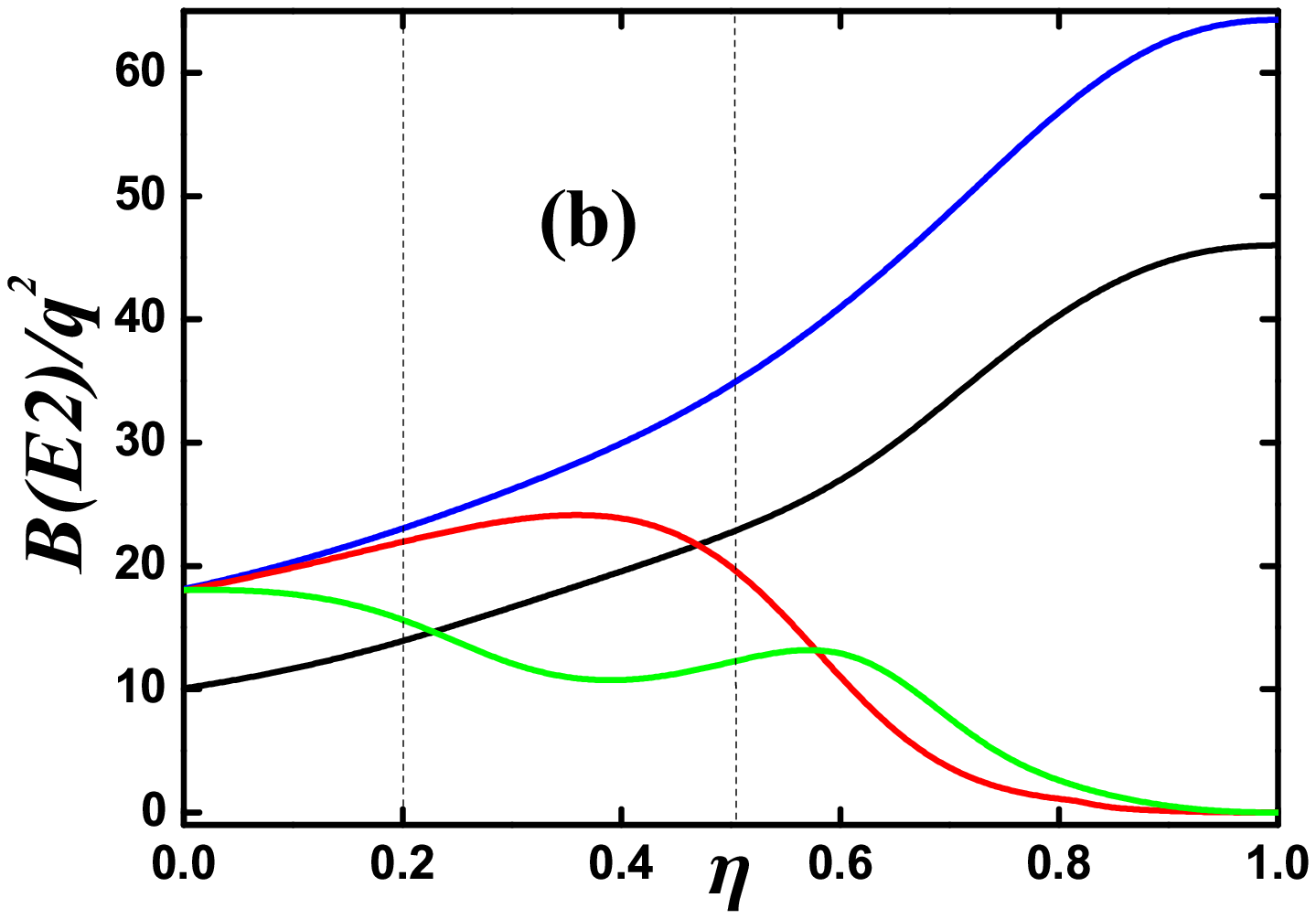}
\caption{(a) The evolutional behaviors of the partial low-lying levels as a function of $\eta$ for $N=10$ along the real red line in Fig. 1(b); (b) The evolutional behaviors of the $B(E2; 2_{1}^{+}\rightarrow 0_{1}^{+})$ (black line), $B(E2; 4_{1}^{+}\rightarrow 2_{1}^{+})$ (blue line), $B(E2; 2_{2}^{+}\rightarrow 2_{1}^{+})$ (red line), $B(E2; 0_{2}^{+}\rightarrow 2_{1}^{+})$ (green line) along the real red line in Fig. 1(b).}
\end{figure}

Fig. 2(a) also shows the energy evolution of the ground state of $\hat{H}$ along the real red line in Fig. 1(b) for $N=10$ (dashed red line) and for $N=60$ (real red line). For $N=60$, it is shown that, around $\eta=0.5$ (denoted by the middle dashed line), a new shape phase transition appears. The part of the real red line deviating from the real blue line is prominent, which has a steeper descent. Between $\eta=0$ and $\eta=0.2$, there is the spherical shape, and between $\eta=0.5$ and $\eta=1$, there is the prolate shape. Obviously, between the two shapes, the new $\gamma$-softness exists. When $\kappa$ changes from $\kappa_{0}$ to $0.9\kappa_{0}$, the new $\gamma$-soft region reduces, but it does exist. A key point is that, between $\eta=0.2$ and $\eta=0.5$, the red and blue lines are nearly degenerate, so when $\kappa$ changes from $\kappa_{0}$ to $0.9\kappa_{0}$, the energies of the new $\gamma$-softness are nearly the same and do not reduce. This implies that, the new $\gamma$-softness is really a phase shape. Thus along the left real red line, double shape quantum phase transition can occur. This can not occur for the shape phase transition along the real blue line and in previous IBM.

Some details need further elaboration. For $N=10$, the real blue line in Fig. 1(b) is nearly the critical line between the prolate shape and the new $\gamma$-soft region, so at the left side of the real blue line, it seems only the prolate shape region. This implies that the new $\gamma$-soft region is larger than expected. which will be further clarified in \cite{Zhao2}. For $N=35$, the critical line curves to the left, so the new shape phase transition becomes more prominent.

\begin{figure}[tbh]
\includegraphics[scale=0.33]{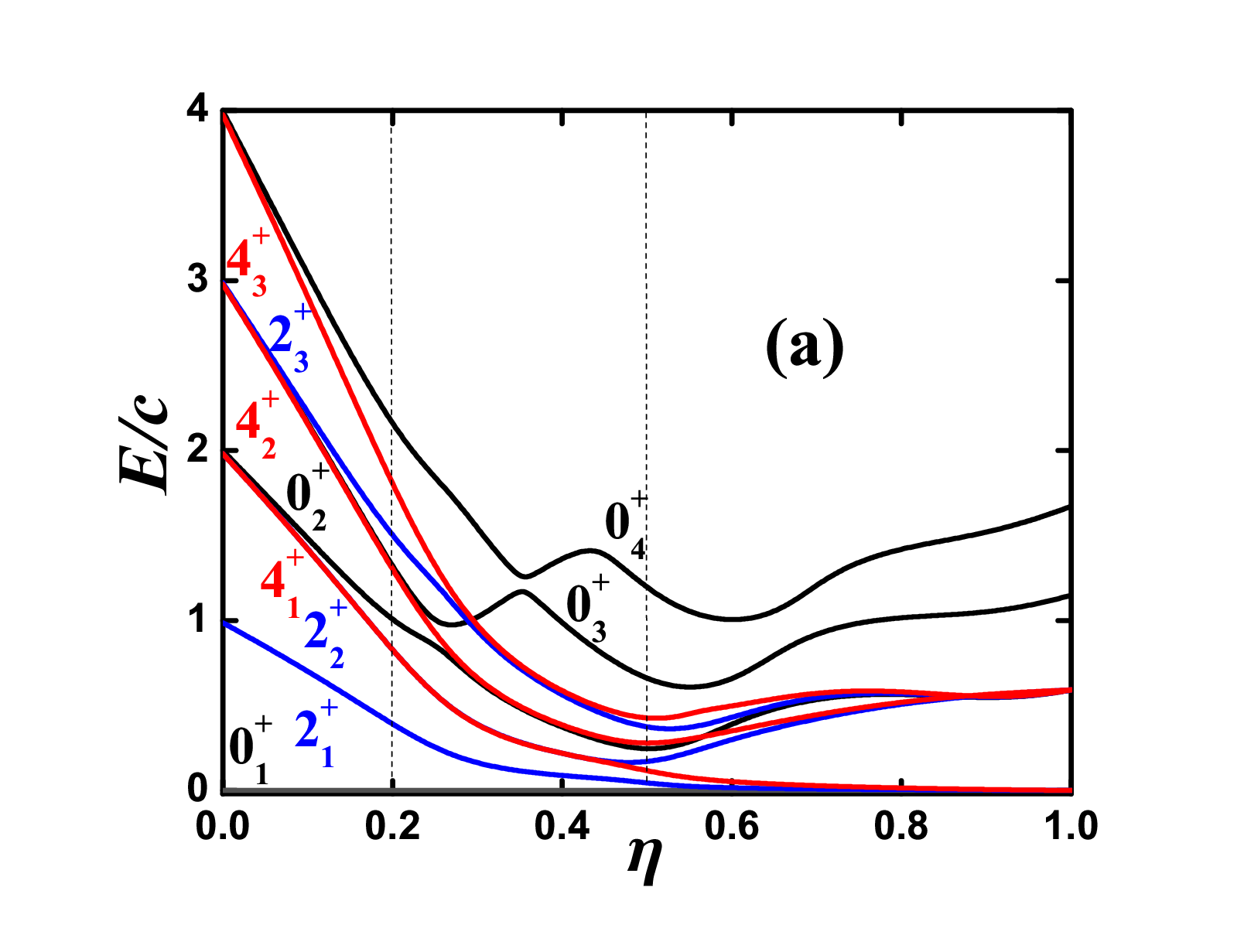}
\includegraphics[scale=0.33]{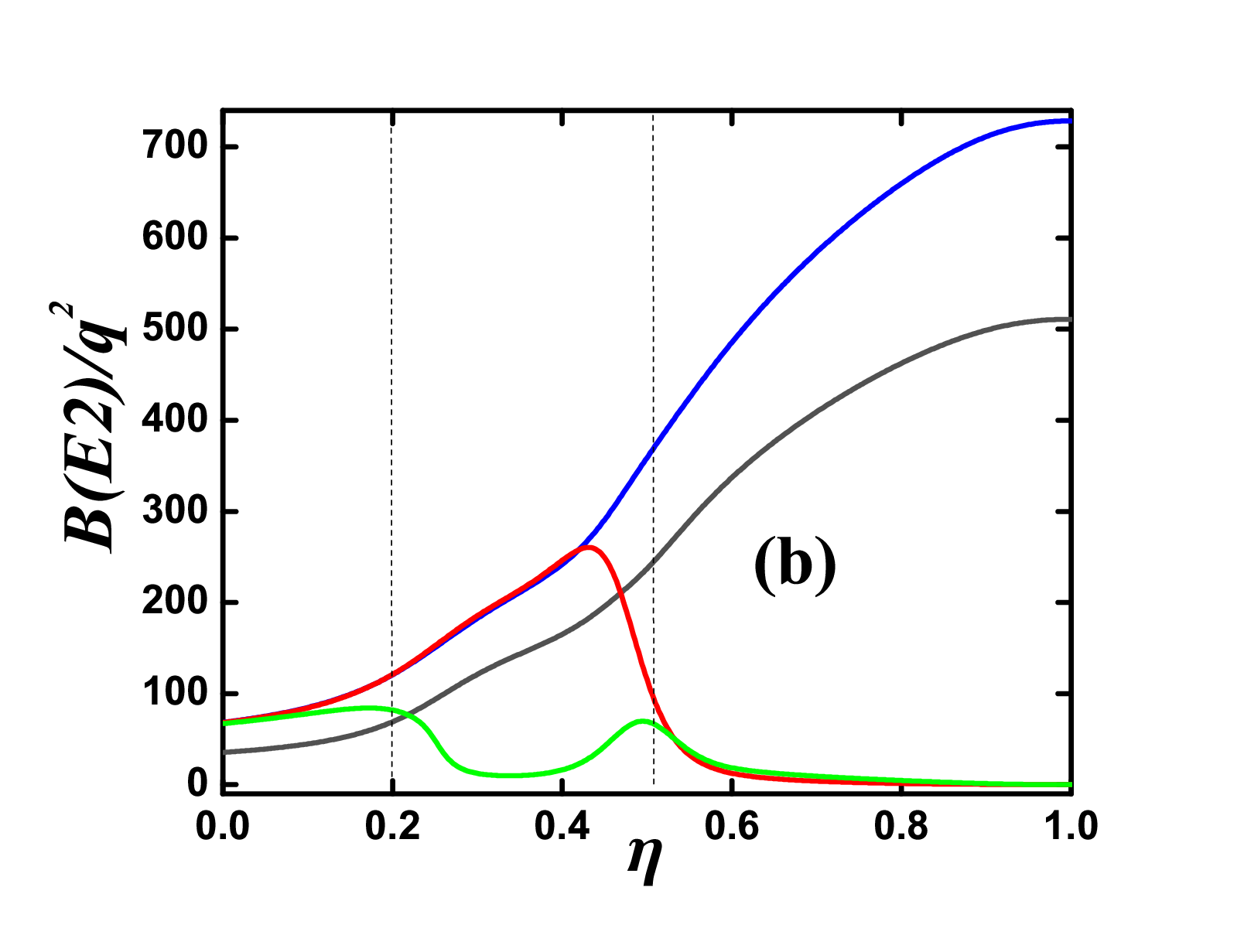}
\caption{(a) The evolutional behaviors of the partial low-lying levels as a function of $\eta$ for $N=35$ along the real red line in Fig. 1(b); (b) The evolutional behaviors of the $B(E2; 2_{1}^{+}\rightarrow 0_{1}^{+})$ (black line), $B(E2; 4_{1}^{+}\rightarrow 2_{1}^{+})$ (blue line), $B(E2; 2_{2}^{+}\rightarrow 2_{1}^{+})$ (red line), $B(E2; 0_{2}^{+}\rightarrow 2_{1}^{+})$ (green line) along the real red line in Fig. 1(b).}
\end{figure}

In previous IBM, similar result of the ground energy evolution along the real blue line can be also obtained along the evolution path from the U(5) symmetry limit to the O(6) symmetry limit. If the O(6) symmetry limit is the prolate-oblate critical point, the connected line in the deformed region between the U(5) symmetry limit and the O(6) symmetry limit is a prolate-oblate critical line. When deviating from the critical line, the energies of the deformed region reduce and the double shape phase transitions can not be observed \cite{Wang08}.

\begin{figure}[tbh]
\includegraphics[scale=0.33]{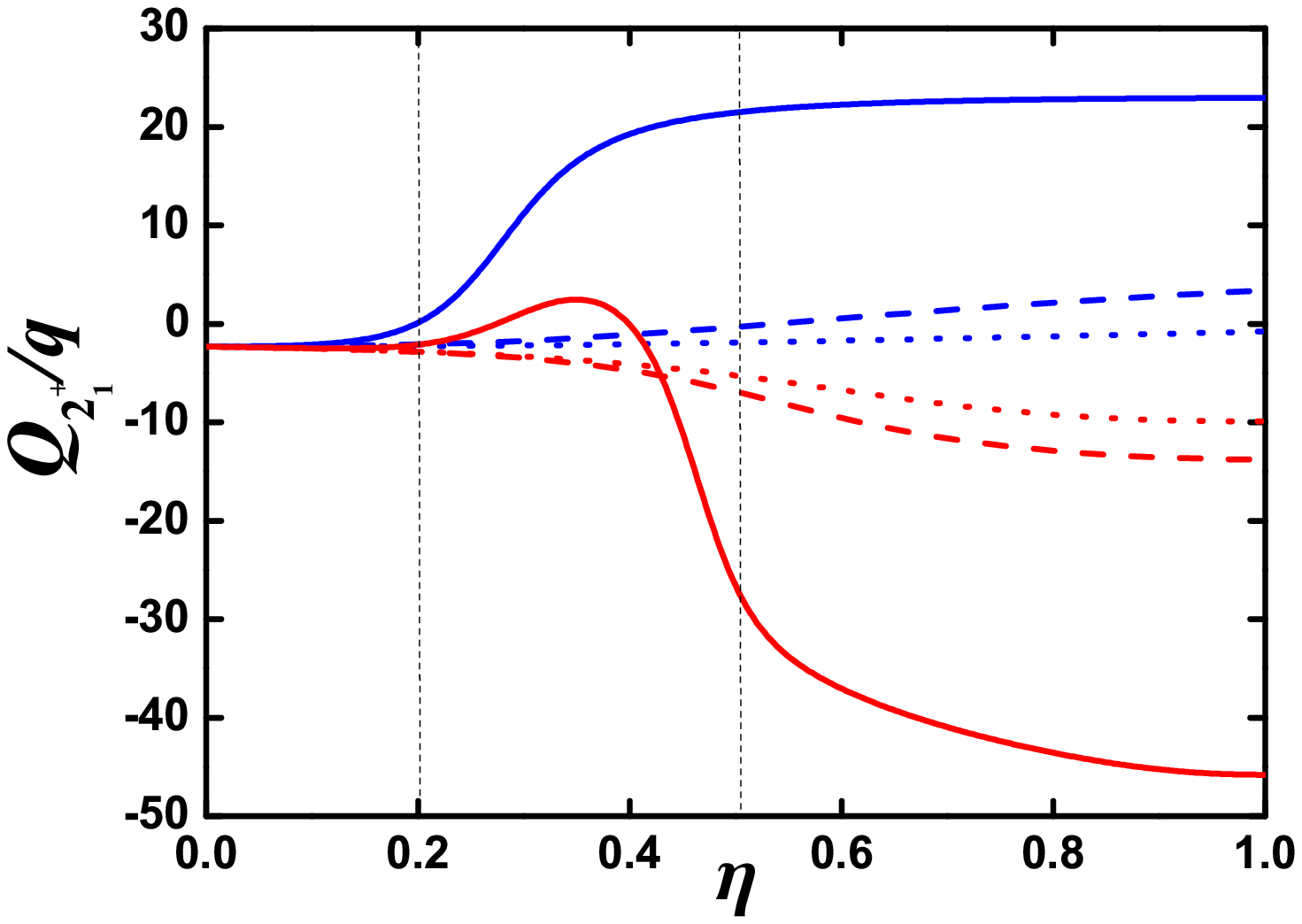}
\caption{The $Q_{2_{1}^{+}}$ evolution of the $2_{1}^{+}$ state of $\hat{H}$ along the real blue (red) line in Fig. 1(b) for $N=7$ (dotted blue (red) line), $N=10$ (dashed blue (red) line) and for $N=35$ (real blue (red) line).}
\end{figure}

The mean value of the $d$ boson number in the ground state $\bar{n}_{d}$ is also important. Fig. 2(b) shows the $\bar{n}_{d}$ evolution along the real blue line in Fig. 1(b) for $N=10$ (dashed blue line) and $N=60$ (real blue line). The phase transition behaviors from the spherical to the new $\gamma$-soft across $\eta=0.2$ is clear. Fig. 2(b) also shows the $\bar{n}_{d}$ evolution along the real red line in Fig. 1(b) for $N=10$ (dashed red line) and $N=60$ (real red line). The double shape quantum phase transitions are also clear. When $\eta>0.5$, the two red lines are deviated from the two blue lines obviously. And importantly between $\eta=0.2$ and $\eta=0.5$ the red and blue lines are nearly degenerate for $N=10$ and $N=60$. The new $\gamma$-soft phase really exists.

Now we discuss some observable quantities, such as the excited energies, the B(E2) values, and the electric quadrupole moment of the $2_{1}^{+}$ state. Previous discussions can help us confirm that the double shape quantum phase transitions do exist. These observable quantities can help us find them experimentally.

We first study the shape phase transition along the real blue line in Fig. 1(b) from the U(5) symmetry limit to the SU(3) degenerate point. This study has been performed in \cite{Wang22} for $N=7$. Here the evolutional behaviors of the partial low-lying states for $N=10$ are shown in Fig. 3(a). The $4_{1}^{+}$ and $2_{2}^{+}$ states are nearly degenerate. $\eta=0.5$ presents the spherical-like spectra, in which the energy of the $0_{3}^{+}$ state is nearly twice the one of the $0_{2}^{+}$ state. The spherical-like spectra was confirmed in $^{106}$Pd recently \cite{WangPd}. Thus the shape phase transition discussed in this paper can be found in Pd nuclei. Fig. 3(b) shows the evolutional behaviors of the B(E2) values of the $B(E2; 2_{1}^{+}\rightarrow 0_{1}^{+})$, $B(E2; 4_{1}^{+}\rightarrow 2_{1}^{+})$, $B(E2; 2_{2}^{+}\rightarrow 2_{1}^{+})$, $B(E2; 0_{2}^{+}\rightarrow 2_{1}^{+})$ along the real blue line in Fig. 1(b). The results are similar to the evolutions from the U(5) symmetry limit to the O(6) symmetry limit.

\begin{figure}[tbh]
\includegraphics[scale=0.33]{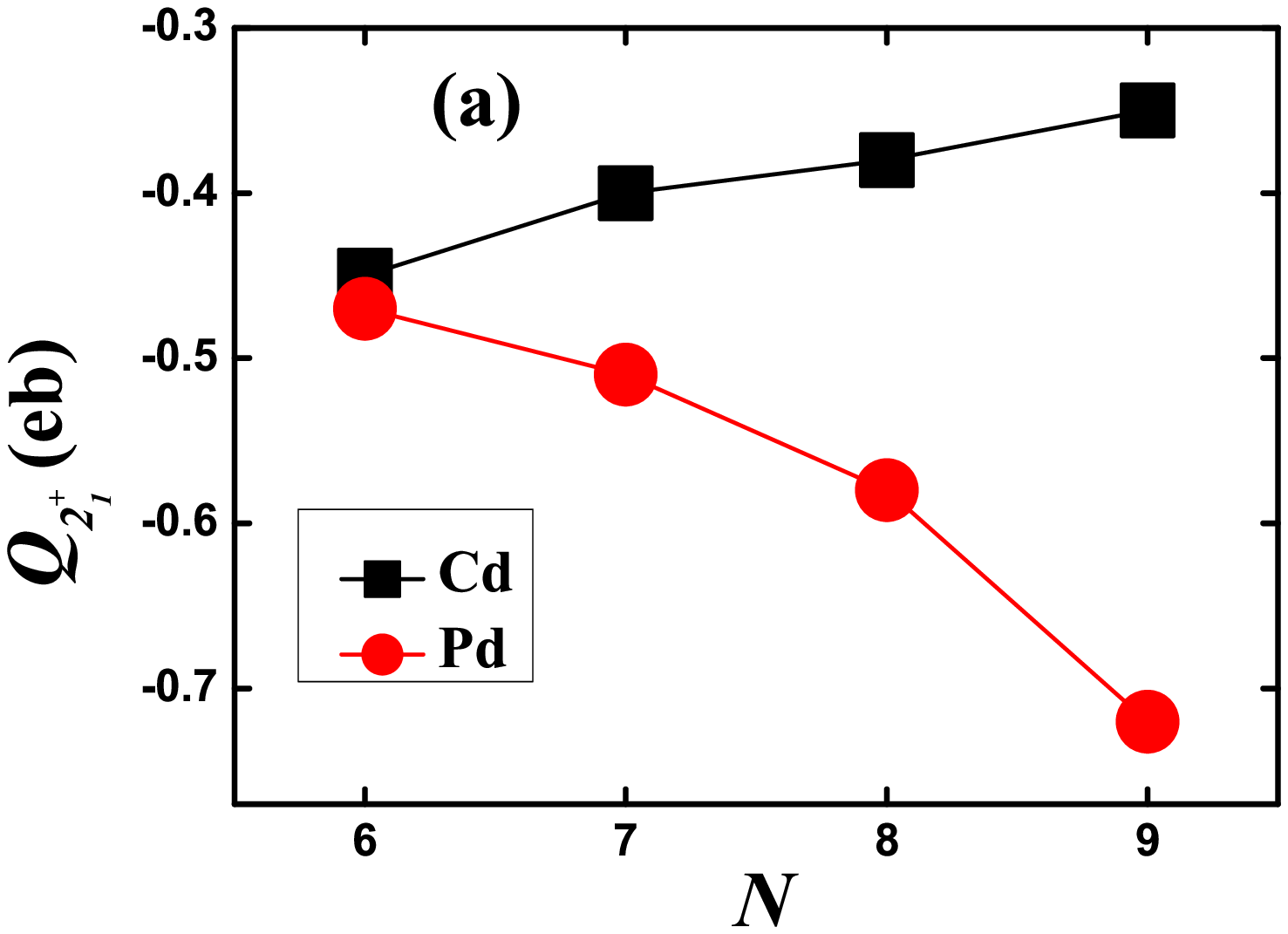}
\includegraphics[scale=0.33]{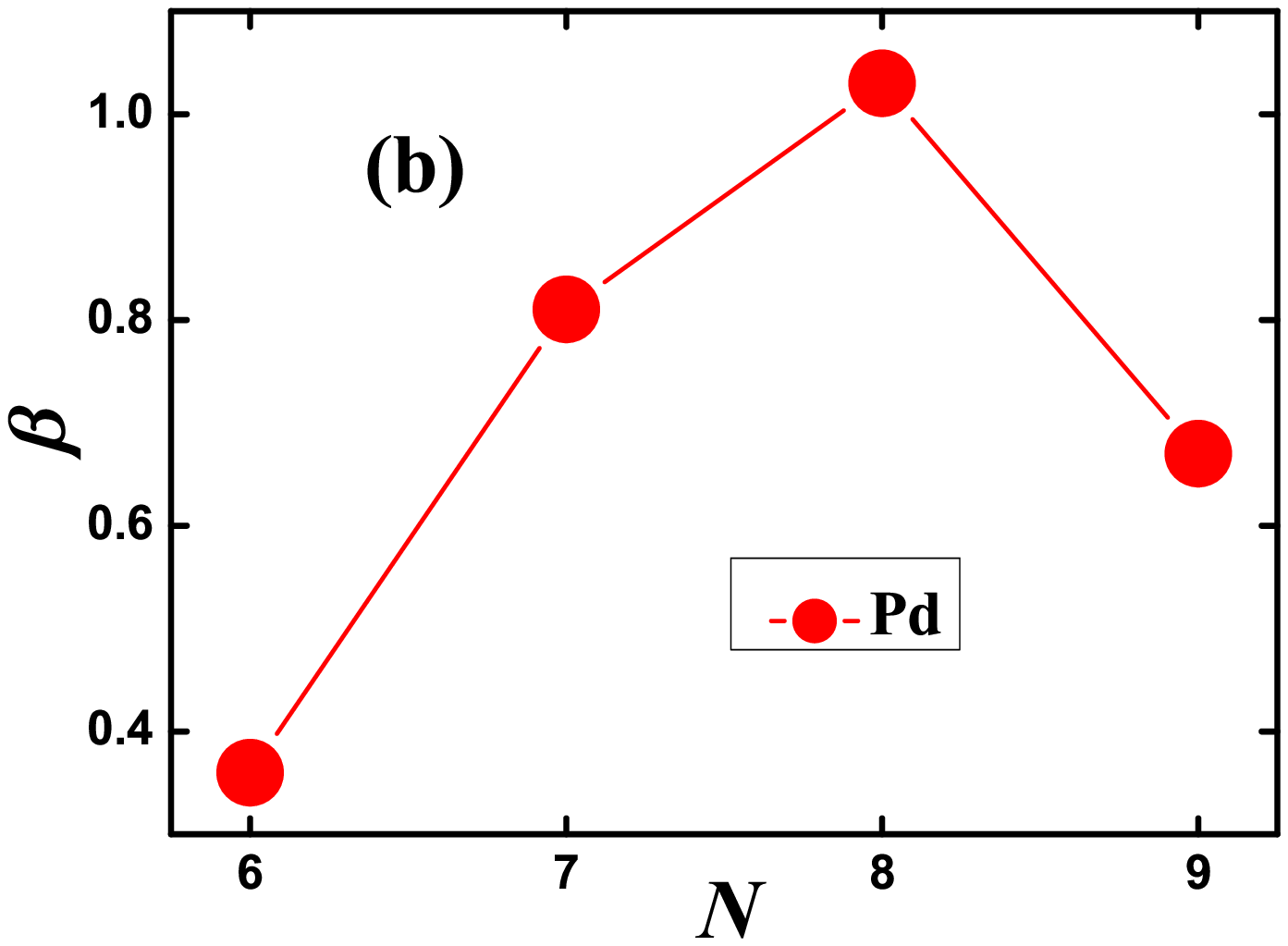}
\caption{(a) Different evolutional trends of the $Q_{2_{1}^{+}}$ values of $^{108-114}$Cd and $^{104-110}$Pd; (b) Evolutional behavior of the $\beta$ values of $^{104-110}$Pd.}
\end{figure}

Fig. 4(a) presents the evolutional behaviors of the partial low-lying states for $N=35$ along the real blue line in Fig. 1(b), which is a new result. The shape phase transition from the spherical shape to the new $\gamma$-soft rotation becomes more prominent. When $\eta>0.2$, the $4_{1}^{+}$ and $2_{2}^{+}$ states begin to separate because the degenerate line curves to the left. Besides, the level-anticrossing of the $0_{2}^{+}$, $0_{3}^{+}$, $0_{4}^{+}$ states becomes more clear \cite{Li25,Jolie02,Arias03}. Fig. 4(b) shows the evolutional behaviors of the B(E2) values of the $B(E2; 2_{1}^{+}\rightarrow 0_{1}^{+})$, $B(E2; 4_{1}^{+}\rightarrow 2_{1}^{+})$, $B(E2; 2_{2}^{+}\rightarrow 2_{1}^{+})$, $B(E2; 0_{2}^{+}\rightarrow 2_{1}^{+})$ along the real blue line in Fig. 1(b). The shape phase transition becomes more clear. When $N$ increases, the values of $B(E2; 2_{2}^{+}\rightarrow 2_{1}^{+})$, $B(E2; 0_{2}^{+}\rightarrow 2_{1}^{+})$ becomes smaller. 

Now we discuss the double shape phase transitions along the real red line in Fig. 1(b). Fig. 5(a) presents the evolutional behaviors of the partial low-lying states for $N=10$. Between the $\eta=0.2$ and $\eta=0.5$, the $4_{1}^{+}$ and $2_{2}^{+}$ are nearly degenerate, and the spectra are also similar to the spherical-like spectra, so this is the new $\gamma$-soft phase. When $\eta>0.5$, obviously it is the prolate shape. In Fig. 6(a) the case of $N=35$ is shown and the shape phase transition from the $\gamma$-soft rotation to the prolate shape becomes very prominent. Thus $k=0.9\kappa_{0}$ and $\eta=0.5$ is the phase transition critical point.

In previous IBM, the O(6) $\gamma$-softness is the shape phase transition critical point from the prolate shape to the oblate shape, so it is not a shape phase. There is no shape phase transition from the O(6) symmetry limit ($\gamma$-soft rotation) to the SU(3) symmetry limit (prolate shape). In the SU3-IBM, the new $\gamma$-softness is a shape phase and the shape phase transition from the new $\gamma$-soft phase to the prolate shape really exists.

Fig. 5(b) and Fig. 6(b) present the evolutional behaviors of the B(E2) values of the $B(E2; 2_{1}^{+}\rightarrow 0_{1}^{+})$, $B(E2; 4_{1}^{+}\rightarrow 2_{1}^{+})$, $B(E2; 2_{2}^{+}\rightarrow 2_{1}^{+})$, $B(E2; 0_{2}^{+}\rightarrow 2_{1}^{+})$ for $N=10$ and $N=35$.  The double shape phase transitions are also clear. Across $\eta=0.5$, the value of $B(E2; 0_{2}^{+}\rightarrow 2_{1}^{+})$ increases first, and then decreases, which can not be found in Fig. 3(b) and Fig. 4(b).

In \cite{Wang25}, the first evidence confirming the existence of a sphere-like spectra was found, which is the anomalous evolutional trend of the electric quadrupole moments of the first $2_{1}^{+}$ states $Q_{2_{1}^{+}}$ in Cd nuclei. Now we further study this interesting phenomenon. Fig. 7 shows the evolutional behaviors of the  $Q_{2_{1}^{+}}$ values along the real blue line (blue lines in Fig. 7) and the real red line (red lines in Fig. 7) for $N=7$ (dotted), $N=10$ (dashed) and $N=15$ (real). Along the real red line in Fig. 1(b), the double shape phase transitions can be clearly observed. The key result is the different evolutional trends of the two parameter paths. The blue one is anomalous. When $N$ increases, the value evolves to the oblate side. The red line is just opposite. When $N$ increases, its magnitude increases too if $\eta\geq 0.5$.

Finally, we look for some experimental evidences for the existence of the shape phase transition from the new $\gamma$-soft phase to the prolate shape. Fig. 8(a) shows the different evolutional trends of the $Q_{2_{1}^{+}}$ values of $^{108-114}$Cd and $^{104-110}$Pd. This was first observed in \cite{Wang25}, and can be regarded as the first strong support for the existence of the spherical-like spectra. The discussions in this paper further support this conclusion and is an indirect experimental support for the existence of the shape phase transition. A direct support can be observed in Fig. 8(b). In the previous analysis, we see that, from the new $\gamma$-soft phase to the prolate shape, the value of $B(E2; 0_{2}^{+}\rightarrow 2_{1}^{+})$ increases first, and then decreases. here we define $\beta=B(E2; 0_{2}^{+}\rightarrow 2_{1}^{+})/B(E2; 2_{1}^{+}\rightarrow 0_{1}^{+})$. Fig. 8(b) presents the $\beta$ evolution of $^{104-110}$Pd and obviously it is clearly in line with the theoretical prediction. Thus $^{104,106}$Pd are two typical new $\gamma$-soft nuclei \cite{WangPd} and $^{108}$Pd may be a critical nucleus from the new $\gamma$-soft phase to the prolate shape. However for finite-$N$, the spherical-like spectra and the critical spectra may be not distinguished. A detailed discussions on the properties of $^{104-110}$Pd will be given in a future paper.

It should be noticed that this double shape quantum phase transitions cannot be verified directly. In the last two decades, the experimental discovery that nuclei previously thought to be spherical cannot be confirmed to be spherical is a breakthrough in the field of nuclear structure \cite{Garrett18}. If the spherical nucleus is absent, it is difficult to confirm the shape phase transition from the spherical shape to the new $\gamma$-soft phase. In our discussion, we also found that the spherical nucleus and the critical nucleus at $\eta=0.2$ are also difficult to be distinguished, which can help us further discuss those nuclei that look like spherical.

\section{Conclusion}

Based on the existence of the sphere-like spectra \cite{Wang22,Wang25,WangPd}, we further discuss the related shape quantum phase transition. In this paper, we have drawn some new conclusions. First, the new $\gamma$-softness is a shape phase, which is very different from previous O(6)-softness as a critical point. Then, we find the double quantum phase transitions along a sing parameter path. We confirm that there is indeed a shape phase transition from the new $\gamma$-soft phase to the prolate shape, and we find experimental evidence that $^{108}$Pd may be a critical nucleus, which will be studied in detail later. In next paper, the scope of the new $\gamma$-soft region in the SU3-IBM for finite-$N$ will be given and the oblate side is also discussed \cite{Zhao2}.

\end{document}